\documentclass[preprint,aps,superscriptaddress,nofootinbib]{revtex4}

\usepackage{epsf}

\newcommand{\beqa}{\begin{eqnarray}}
\newcommand{\eeqa}{\end{eqnarray}}
\newcommand{\beq}{\begin{equation}}
\newcommand{\eeq}{\end{equation}}

\def\d{{\rm d}}

\def\lqcd{\Lambda_{\rm QCD}}
\def\ksla{k\!\!\!\slash}
\def\psla{p\hspace{-5pt}\slash}
\def\qsla{q\hspace{-5.5pt}\slash} 
\def\vsla{v\hspace{-5.5pt}\slash}

\begin{document}

\preprint{\vbox{\hbox{FERMILAB-Pub-02/079-T} \hbox{LBNL--50242} 
  \hbox{CALT--68--2384} \hbox{hep-ph/0205148} }}

\vspace*{1.5cm}

\title{\boldmath Enhanced subleading structure functions in semileptonic $B$
decay}

\author{Adam K.\ Leibovich}\email{adam@fnal.gov}
\affiliation{Theory Group, Fermi National Accelerator Laboratory,
	Batavia IL 60510\vspace{6pt} }

\author{Zoltan Ligeti}\email{zligeti@lbl.gov}
\affiliation{Ernest Orlando Lawrence Berkeley National Laboratory,
        University of California, Berkeley CA 94720\vspace{6pt} }

\author{Mark B.\ Wise\,}\email{wise@theory.caltech.edu}
\affiliation{California Institute of Technology, Pasadena, CA 91125
	\\[20pt] $\phantom{}$ }

\begin{abstract} \vspace*{8pt}

The charged lepton spectrum in semileptonic $B\to X_u \ell\bar\nu$ decay near
maximal lepton energy receives important corrections from subleading structure
functions that are formally suppressed by powers of $\lqcd/m_b$ but are
enhanced by numerical factors.  We investigate the series of higher order terms
which smear over a region of width $\Delta E_\ell \sim \lqcd$ near the endpoint
the contributions proportional to $\delta(E_\ell - m_b/2)$ times (i)~the matrix
element of the chromomagnetic operator, and (ii)~four-quark operators.  These
contribute to the total rate at the few percent level, but affect the endpoint
region much more significantly.  Implications for the determination of
$|V_{ub}|$ are discussed.

\end{abstract}

\maketitle

\section{Introduction}

Extracting the CKM matrix element $|V_{ub}|$ from the inclusive decay $B\to
X_u\ell\bar\nu$ is complicated due to the large background from the decay $B\to
X_c\ell\bar\nu$.  A way to remove this background is to impose a lower cut on
the charged lepton energy~\cite{oldVub}, restricting it to be within a few
hundred MeV of the kinematic endpoint. Unfortunately, in this region of phase
space there are large corrections to the decay rate due to the $b$ quark
distribution function in the $B$ meson, which introduces model dependence into
the extraction of $|V_{ub}|$.  This uncertainty can be reduced by using the
$B\to X_s\gamma$ photon spectrum as an input \cite{shape,extractshape},
which was recently carried out by the CLEO collaboration~\cite{cleoVub}.

The structure function which describes the lepton endpoint region in $B\to
X_u\ell\bar\nu$ and the photon endpoint region in $B\to X_s\gamma$ arises due
to an infinite series of formally higher order terms in the operator product
expansion (OPE) becoming leading in the endpoint region.  Corrections to this
description in the endpoint region are only suppressed by one power of
$\lqcd/m_b$ (although nonperturbative effects in the total rate are suppressed
by $\lqcd^2/m_b^2$), but can be enhanced by other factors, as explained below. 
We concentrate on corrections which describe the smearing of certain terms in
the OPE of order $(\lqcd/m_b)^2\, \delta(1-y)$ and $(\lqcd/m_b)^3\,
\delta(1-y)$, where $y = 2E_\ell/m_b$.  There are important terms at such
orders, enhanced by numerical factors, which contribute to the $B\to
X_u\ell\bar\nu$ rate at the several percent level.  Since these contributions
are concentrated near the endpoint, their effects on the rate in a restricted
region can be an order of magnitude larger.  

The $B\to X_u \ell\bar\nu$ lepton spectrum, including dimension-5
operators~\cite{OPE} and neglecting perturbative corrections, is given by
\beqa\label{slspec}
{\d \Gamma\over \d y} &=& {G_F^2\, m_b^5\, |V_{ub}|^2\, \over 192\,\pi^3}\,
\bigg\{ 2y^2(3-2y)\, \theta(1-y)
  - {2\lambda_2\over m_b^2} \bigg[ \frac{11}2\, \delta(1-y) 
  - y^2(6+5y)\, \theta(1-y) \bigg] \nonumber\\
&&\qquad\qquad\qquad - {2\lambda_1\over m_b^2} \bigg[ \frac16\, \delta'(1-y) 
  + \frac16\, \delta(1-y) - \frac53\, y^3\, \theta(1-y) \bigg] + \ldots 
  \bigg\} ,
\eeqa
where
\beq\label{l12def}
\lambda_1 = \frac12 \langle B(v)|\, \bar b_v\, (iD)^2\, b_v\, |B(v)\rangle\,, 
  \qquad
\lambda_2 = \frac16 \langle B(v)|\, \bar b_v\, {g_s\over2}\, 
  \sigma_{\mu\nu}\, G^{\mu\nu}\, b_v |B(v)\rangle\,,
\eeq
and $b_v$ denotes the $b$ quark field in HQET.  While $\lambda_1$ is not known
precisely, the $B^*-B$ mass difference implies that $\lambda_2 = 0.12\,
\mbox{GeV}^2$. Of the terms in Eq.~(\ref{slspec}), the leading order (in
$\lqcd/m_b$) structure function contains $2 [\theta(1-y) - \lambda_1/(6m_b^2)\,
\delta'(1-y) + \ldots]$.  The derivative of the same combination occurs in the
$B\to X_s\gamma$ photon spectrum~\cite{FLS}, given by
\begin{equation}\label{bsgspec} 
{\d \Gamma\over \d x} = 
  {G_F^2\, m_b^5 \,|V_{tb}V_{ts}^*|^2\, \alpha\, C_7^2 \over32\,\pi^4}\, 
  \bigg[ \bigg( 1 + {\lambda_1-9\lambda_2\over2m_b^2} \bigg)\, \delta(1-x) 
  - {\lambda_1 + 3\lambda_2 \over 2m_b^2}\, \delta'(1-x) 
  - {\lambda_1 \over 6m_b^2}\, \delta''(1-x) + \ldots \bigg] ,
\end{equation} 
where $x = 2E_\gamma/m_b$.  The effect of the leading order structure function
on the semileptonic decay is reviewed in Sec.~II. At subleading order,
proportional to $\delta(1-y)$ in Eq.~(\ref{slspec}) and to $\delta'(1-x)$ in
Eq.~(\ref{bsgspec}), possibly the most striking difference is between the terms
involving $\lambda_2$, whose coefficient is $11/2$ in Eq.~(\ref{slspec}) and
$3/2$ in Eq.~(\ref{bsgspec}).  Because of the 11/2 factor, the  $\lambda_2\,
\delta(1-y)$ term is important in the lepton endpoint region and it contributes
about $-5.5\%$ to the total $B\to X_u \ell\bar\nu$ rate.  This contribution is
investigated in Sec.~III.

At order $\lqcd^3/m_b^3$, there are dimension-6 four-quark operators in the
OPE,
\beq\label{dim6ops}
O_{V-A} = (\bar b_v \gamma^\mu P_L u)\, (\bar u \gamma_\mu P_L b_v) , \qquad
O_{S-P} = (\bar b_v P_L u)\, (\bar u P_L b_v) ,
\eeq
where $P_L = (1-\gamma_5)/2$.  Compared to the lower dimensional operators
discussed so far, these are enhanced by $16\pi^2$ and also enter the $B\to
X_u\ell\bar\nu$ electron spectrum~\cite{Voloshin} as delta function
contributions at the endpoint
\begin{equation}\label{volo}
{\d \Gamma_{(6)}\over \d y} = - \frac{G_F^2\,  m_b^2\, |V_{ub}|^2}{12\pi}\, 
  f_B^2\, m_B\, (B_1 - B_2)\, \delta(1-y) \,.
\end{equation}
Here $f_B$ is the $B$ meson decay constant and $B_{1,2}$ parameterize the
matrix elements of the operators in Eq.~(\ref{dim6ops}) between $B$ meson
states,
\beq\label{bagparameters}
\frac12 \langle B | O_{V-A} | B \rangle = \frac{f_B^2\, m_B}8\, B_1 , \qquad
\frac12 \langle B | O_{S-P} | B \rangle = \frac{f_B^2\, m_B}8\, B_2 \,.
\eeq
If one assumed factorization and used the vacuum saturation approximation
then these contribution would vanish, since $B_1 = B_2 = 1\ (0)$ for charged
(neutral) $B$ mesons.  There is no reason to believe that factorization holds
exactly, and deviations from it at the 10\% level are quite
possible~\cite{Voloshin}.  With such an estimate (i.e., $|B_1 - B_2| = 0.1$,
and $f_B = 200\,$MeV), the four-quark operators contribute of order 3\% to the
total $B\to X_u \ell\bar\nu$ rate. This contribution, which is absent from the
$B\to X_s\gamma$ rate, is investigated in Sec.~IV.

Section~V contains our conclusions and the implications for the determination
of $|V_{ub}|$.

\section{The leading order structure function}

Near the endpoint region of the $B\to X_u\ell\bar\nu$ lepton energy spectrum,
the OPE is effectively an expansion in $\lqcd/[m_b(1-y)]$ instead of
$\lqcd/m_b$.  Thus, within $\Delta E_\ell \sim \lqcd$ from the endpoint, there
is an infinite series of terms of the form
\beq\label{leadspec}
{\d\Gamma\over \d y} = {G_F^2 |V_{ub}|^2\, m_b^5\over 192\,\pi^3}\, 
  \sum_{n=0}^\infty {A_n\over m_b^n\, n!}\, 2\,\theta^{(n)}(1-y) \,,
\eeq
which are equally important when smeared over $\Delta y \sim
\lqcd/m_b$~\cite{shape}.  Here $A_n$ are given by the matrix elements
\beq
\frac12 \langle B(v) |\, \bar b_v\, i D_{\alpha_1} \ldots 
  iD_{\alpha_n}\, b_v | B(v)\rangle 
= A_n\, v_{\alpha_1} \ldots v_{\alpha_n} + \mbox{(other tensor structures)}\,,
\eeq
and satisfy $A_0 = 1$, $A_1 = 0$, $A_2 = -\lambda_1/3$, etc.  These
terms can be formally resummed into a nonperturbative structure function for
the $b$ quark,
\beq\label{leadstruc}
f(k_+) = \frac12 \langle B(v) |\, \bar b_v\, \delta(k_+ - iD \cdot n)\, b_v 
  | B(v)\rangle \,,
\eeq
where $n$ is a light-like vector satisfying $n \cdot v = 1$ and $n^2=0$.  The
moments of $f(k_+)$ satisfy $\int \d k_+\, f(k_+)\, k_+^n = A_n$.  This is
convenient for model calculations because the effects of the distribution
function can be included by replacing $m_b$ by $m_b^* \equiv m_b + k_+$, and
integrating over $k_+$,
\beq\label{smear}
{{\rm d}\Gamma\over {\rm d}E_\ell} = \int {\rm d}k_+\, f(k_+)\,
  {{\rm d}\Gamma_{\rm p}\over {\rm d}E_\ell} \Bigg|_{m_b\to m_b^*} ,
\eeq
where ${\rm d}\Gamma_{\rm p}/{\rm d}E_\ell$ is the parton-level spectrum.  In
the region we consider in this paper, $E_\ell > 2\,$GeV, taking the parton
level spectrum as $2y^2(3-2y)\, \theta(1-y)$ or just the leading twist part of
it, $2\, \theta(1-y)$, makes a negligibly small difference numerically, and
therefore we will use the latter.  (Perturbative order $\alpha_s$ corrections
are also neglected.)

For purposes of illustration, we will use a simple model for the structure
function which has three parameters,
\beq\label{shapemodel}
f(k_+) = {1\over \bar\Lambda}\, {a^{a b} \over \Gamma(a b)}\,
 (1-x)^{a b-1}\, e^{-a(1-x)}\, \theta(1-x) \,, \qquad
x = {k_+\over \bar\Lambda}\,.
\eeq
The parameterization of Ref.~\cite{models2} corresponds to $b \to 1$ and $a \to
a+1$.  The $b$ parameter is introduced because the first moment of the
subleading structure function that occurs in Sec.~IV need not vanish, contrary
to the leading order structure function in Eq.~(\ref{leadstruc}) and the
subleading structure function that follows from the one-gluon matrix element
discussed in Sec.~III.  The first three moments of $f(k_+)$ are given by
\beqa
&& \int \d k_+\, f(k_+) = 1 \,, \qquad
  \int \d k_+\, f(k_+)\,k_+ = \bar\Lambda\, (1-b) \,, \nonumber\\*
&& \int \d k_+\, f(k_+)\,k_+^2
  = \bar\Lambda^2\, \bigg[ (1-b)^2 + {b\over a} \bigg] \,.
\eeqa
For the leading order structure function we will use Eq.~(\ref{shapemodel})
with $b = 1$, $\bar\Lambda = 0.5\,$GeV, and $\lambda_1 = -3 \bar\Lambda^2 / a =
-0.3\,\mbox{GeV}^2$, which are in the ballpark of the recent CLEO
determinations~\cite{cleomom}.

\begin{table}[t]
\begin{tabular}{@{\hspace{10pt}} c @{\hspace{10pt}} || 
  @{\hspace{10pt}} c @{\hspace{10pt}} |
  @{\hspace{10pt}} c @{\hspace{10pt}} c @{\hspace{10pt}} c @{\hspace{10pt}}}
  \hline\hline
Lepton energy  &  \multicolumn{4}{c@{\hspace{10pt}}}{Fraction of events in the
  endpoint region} \\
intervals  &  No smearing  
  &  $\lambda_1 = -0.1\,\mbox{GeV}^2$
  &  $\lambda_1 = -0.3\,\mbox{GeV}^2$  
  &  $\lambda_1 = -0.5\,\mbox{GeV}^2$ \\ \hline\hline
$2.0\,\mbox{GeV} < E_\ell$  &  32.6\%  &  32.6\%  &  32.9\%  &  33.4\% \\
$2.1\,\mbox{GeV} < E_\ell$  &  24.3\%  &  24.3\%  &  24.8\%  &  25.7\% \\
$2.2\,\mbox{GeV} < E_\ell$  &  15.9\%  &  16.1\%  &  17.2\%  &  18.3\% \\
$2.3\,\mbox{GeV} < E_\ell$  &  7.5\%   &  8.4\%   &  10.2\%  &  11.6\% \\ 
$2.4\,\mbox{GeV} < E_\ell$  &  n/a     &  2.6\%   &  4.6\%   &  5.9\% \\ 
  \hline\hline
\end{tabular}
\caption{Fractional rate in the endpoint region without and with resummation. 
For the structure function we used Eq.~(\ref{shapemodel}) with $b=1$,
$\bar\Lambda = 0.5\,$GeV, and $\lambda_1$ as shown in the Table.}
\label{tablo}
\end{table}

The effect of resumming the leading singularities is illustrated in
Table~\ref{tablo}.  For the structure function we used Eq.~(\ref{shapemodel})
with $b=1$, $\bar\Lambda = 0.5\,$GeV, and three different values of $\lambda_1
(= -3 \bar\Lambda^2 / a)$.  This Table illustrates that the effect of the
resummation becomes small as the lower limit of the lepton energy interval is
lowered towards 2\,GeV.  Note that we have neglected perturbative corrections
(this is probably the main reason why the numbers in the $\lambda_1 =
-0.3\,\mbox{GeV}^2$ column differ from those in Ref.~\cite{cleoVub}).  The
value of the $\bar\Lambda$ parameter also affects the importance of resumming
the singular terms.  Of course the precise determination of $\bar\Lambda$
(i.e., the $b$ quark mass) is crucial for any extraction of $|V_{ub}|$ from
inclusive $B$ decays.

\section{\boldmath Smearing of the $\lambda_2\, \delta(1-y)$ contributions}

In this section we investigate terms proportional to $\lambda_2\, \delta(1-y)$
in the $B\to X_u \ell\bar\nu$ lepton spectrum (or to $\lambda_2\, \delta'(1-x)$
in the $B\to X_s \gamma$ photon spectrum), which are subleading in the twist
expansion, and therefore do not necessarily drop out from the relation between
the two spectra.  Such corrections originate from three sources (for details
see~\cite{book}):

(i) corrections to the vanishing of the forward scattering matrix element of
any dimension-4 operator, $\langle B(v)|\, \bar b_v\, iD^\alpha\, b_v\, |
B(v)\rangle = 0$,  due to higher order terms in the HQET Lagrangian;

(ii) corrections to the  $b(x) = e^{-i m_b v\cdot x}\, b_v(x)$ relation between
the $b$ quark field in QCD and in HQET;

\begin{figure}[t]
\centerline{\epsfxsize=0.45\textwidth \epsfbox{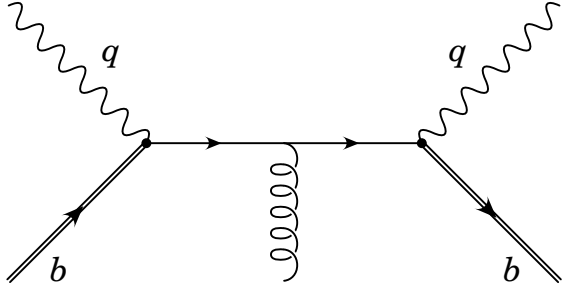}}
\caption{The one-gluon matrix element.}
\label{onegluon}
\end{figure}

(iii) corrections to the leading contribution of the one-gluon matrix element
in Fig.~\ref{onegluon}.

Correspondingly, there are three series of higher dimensional terms in the OPE
that can be resummed into three subleading structure functions that describe
the smearing of these terms in regions of widths $\Delta E_\ell \sim \lqcd$
near the endpoint.  These subleading structure functions that smear the (i),
(ii), and (iii) terms discussed above are denoted in Ref.~\cite{subltwist} by
$t(\omega)$, $h_1(\omega)$, and $h_2(\omega_1,\omega_2)$, respectively.  Since
none of them are known, we will focus on possible effects of only one of them,
which has the largest coefficient.

Of these subleading structure functions, the one related to (i) is universal
and may be absorbed into the redefinition of the leading order structure
function.  Therefore, this function drops out from the relation between the
$B\to X_u \ell\bar\nu$ lepton spectrum and the $B\to X_s \gamma$ photon
spectrum.  To gain insight into the relative importance of the other two
structure functions, note that the coefficient of the $(\lambda_2/m_b^2)\,
\delta'(1-x)$ term in Eq.~(\ref{bsgspec})  arises from the (i)--(iii) terms
respectively as $3/2 = (3 - 2 + 2)/2$~\cite{subltwist}, whereas the
corresponding decomposition of the coefficient of the $(\lambda_2/m_b^2)\,
\delta(1-y)$ term in Eq.~(\ref{slspec}) is $11/2 = (3 + 2 + 6)/2$.  Therefore,
we will focus on the subleading structure function that smears the one-gluon
matrix element (iii).

The one-gluon matrix element in Fig.~\ref{onegluon} is given by the imaginary
part of the $B$ meson matrix element of
\beq
\bar b\, \gamma_\mu \Bigg[ {(m_b \vsla+\ksla-\qsla)\, 
  (i g_s T^A \varepsilon^{A\lambda*}\gamma_\lambda)\, 
  (m_b \vsla+\ksla'-\qsla) \over 
  (m_b v+k-q)^2\, (m_b v+k'-q)^2} \Bigg] \gamma_\nu P_L b \,,
\eeq
where $\varepsilon$ is the gluon polarization vector.  The gluon field strength
arises from the part antisymmetric in the gluon momentum ($p = k - k'$) and
polarization, $p^\mu T^A \varepsilon^{A\nu*} \to -(i/2)G^{\mu\nu}$.  The series
of most important terms in the OPE in the endpoint region is obtained by
expanding in $k$ and $k'$, and keeping the part where in each additional order
in $\lqcd/m_b$ the contribution to the rate becomes more
singular~\cite{shape,Falk:1993vb}.  These terms enter proportional to matrix
elements where each additional covariant derivative gives $v^\alpha$ tensor
structure.  In analogy with $\lambda_2$ in Eq.~(\ref{l12def}), we define
\beqa
&& \sum_{m=0}^n \frac12 \langle B(v) |\, \bar b_v\, i D_{\alpha_1} \ldots 
  i D_{\alpha_m} \bigg[ {g_s\over2}\, \sigma_{\mu\nu}\, G^{\mu\nu} \bigg]
  iD_{\alpha_{m+1}} \ldots iD_{\alpha_n}\, b_v | B(v)\rangle \nonumber\\*
&&\qquad = 3(n+1)\, \lambda_2\, X_n\, v_{\alpha_1} \ldots v_{\alpha_n} 
  + \mbox{(other tensor structures)}\,,
\eeqa
The normalization is chosen such that $X_0 = 1$, and $X_1 = 0$ because of the
heavy quark equation of motion.  The higher order matrix elements are unknown.
The resulting contribution to the $B\to X_u\ell\bar\nu$ lepton spectrum is
\beq\label{1gspec}
{\d\Gamma\over \d y} = -{G_F^2 |V_{ub}|^2\, m_b^3\over 192\,\pi^3}\, 
  6\lambda_2 \sum_{n=0}^\infty {X_n\over m_b^n\, n!}\, \delta^{(n)}(1-y) \,.
\eeq
As for the leading order structure function contribution, the effects of
Eq.~(\ref{1gspec}) can be modelled by smearing the $n=0$ term with a function
$f(k_+)$, as in Eq.~(\ref{smear}), where the moments of $f(k_+)$ must satisfy
$\int \d k_+\, f(k_+)\, k_+^n = X_n$.

\begin{table}[t]
\begin{tabular}{@{\hspace{10pt}} c @{\hspace{10pt}} ||
  @{\hspace{10pt}} c @{\hspace{25pt}} c @{\hspace{20pt}} c @{\hspace{10pt}}}
  \hline\hline
Lepton energy  &  \multicolumn{3}{c@{\hspace{10pt}}}{Fractional correction, 
  $\delta\Gamma/\Gamma$} \\
intervals  &  $a = 2.5$  &  $a = 1$  &  $a = 10$  \\ \hline\hline
$2.0\,\mbox{GeV} < E_\ell$  &  $-$9\%  &  $-$9\%  &  $-$10\% \\
$2.1\,\mbox{GeV} < E_\ell$  &  $-$12\%  &  $-$11\%  &  $-$13\%  \\
$2.2\,\mbox{GeV} < E_\ell$  &  $-$16\%  &  $-$15\%  &  $-$18\%  \\
$2.3\,\mbox{GeV} < E_\ell$  &  $-$24\%  &  $-$23\%  &  $-$27\%  \\
$2.4\,\mbox{GeV} < E_\ell$  &  $-$38\%  &  $-$42\%  &  $-$33\%  \\ \hline\hline
\end{tabular} 
\caption{Fractional corrections to the rate in the endpoint region due to the
$\lambda_2\, \delta(1-y)$ term from the one-gluon matrix element.  The
subleading structure function is modelled by Eq.~(\ref{shapemodel}) with $b=1$,
$\bar\Lambda = 0.5\,$GeV, and the different values of $a$ shown in the Table.  
For the leading structure function $a = 2.5$ is held fixed (corresponding to
$\lambda_1 = -0.3\, \mbox{GeV}^2$), and $b=1$ and $\bar\Lambda = 0.5\,$GeV.}
\label{tab1G}
\end{table}

In Table~\ref{tab1G} we give numerical estimates of the effect of these
corrections on the $B\to X_u\ell\bar\nu$ rate in the endpoint region.  We use
for the leading order structure function the model in Eq.~(\ref{shapemodel})
with $b=1$, $\bar\Lambda = 0.5\,$GeV and $\lambda_1 = -0.3\,\mbox{GeV}^2$. 
Since the leading structure function smears $\theta(1-y)$, the sensitivity to
the precise values of its parameters is relatively smaller than in the case of
the subleading structure functions that smear contributions proportional to
$\delta(1-y)$.  For the subleading structure function we use
Eq.~(\ref{shapemodel}) with $b=1$, $\bar\Lambda = 0.5\,$GeV and with different
values of the $a$ parameter.  Using the $B\to X_s\gamma$
photon spectrum as an input only cancels a third of these corrections, and
furthermore, there is another subleading structure function that contributes
comparably and with the same sign as the one shown in Table~\ref{tab1G}.  The
sign of these corrections is negative in all entries in Table~\ref{tab1G},
because the subleading structure function is assumed to be positive for all
values of $k_+$ and because the contribution of this correction to the total
rate is negative.

\section{\boldmath Smearing of the four-quark operators' contribution}

\begin{figure}[t]
\centerline{\epsfxsize=0.45\textwidth \epsfbox{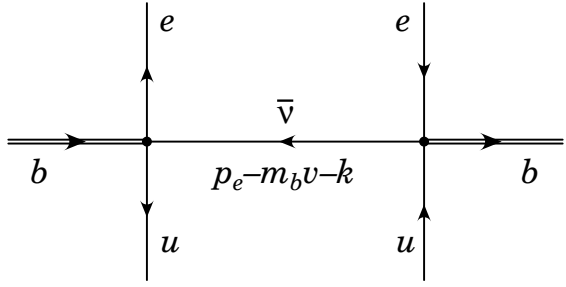}}
\caption{Time ordered product that yields four-quark operators and higher
dimension terms.}
\label{strangeope}
\end{figure}

There are also important contribution to the lepton endpoint region from the
four-quark operators in Eq.~(\ref{dim6ops}).  Compared to the $b$ quark decay
rate, these contributions in the lepton endpoint region are suppressed by
$\lqcd^2/m_b^2$ but are enhanced by $16\pi^2$ (their suppression is 
$\lqcd^3/m_b^3$ in the total rate).  There is again an infinite series of
higher dimensional terms in the OPE which are equally important in the region
of lepton energies within order $\lqcd$ from the endpoint.  These are obtained
from the amplitude in Fig.~\ref{strangeope}, which is proportional to
\beq
(\bar b \gamma_\mu P_L u) \Bigg[ {\bar \ell \gamma^\nu P_L\, 
  i(m_b \vsla + \ksla - \psla_\ell)\, \gamma^\mu P_L \ell \over 
  (m_b v+k-p_\ell)^2} \Bigg] (\bar u \gamma_\nu P_L b) 
\eeq

The series of most singular terms which are equally important in the endpoint
region of the lepton spectrum is obtained by expanding in $k$ and taking the
imaginary part,
\beqa
{\d\Gamma\over \d y} &\propto& \sum_{n=0}^\infty \delta^{(n)}(1-y)\,
  {2^n\over n!\, m_b^n}\, (g^{\mu\nu} - v^\mu v^\nu)\, 
  (v-\hat p_\ell)^{\alpha_1} \ldots (v-\hat p_\ell)^{\alpha_n} \nonumber\\
&&\times  \langle B(v)  | (\bar b_v \gamma_\mu P_L u)\, iD_{\alpha_1} 
  \ldots iD_{\alpha_n} (\bar u \gamma_\nu P_L b_v) | B(v)\rangle \,.
\eeqa
Here $\hat p_\ell = p_\ell/m_b$, and the $(g^{\mu\nu} - v^\mu v^\nu)$ term
arises because in the kinematic region we are considering $v^\mu$ dotted into
the lepton tensor gives a subleading contribution. Since $v-\hat p_\ell$
defines a light-cone direction, $(v-\hat p_\ell)^2$ yields a  suppression by at
least one power of $\lqcd/m_b$.  Therefore, we may write
\beqa
&& \frac12 \langle B(v) | (\bar b_v \gamma_\mu P_L u)\, iD_{\alpha_1} \ldots 
  iD_{\alpha_n} (\bar u \gamma_\nu P_L b_v) | B(v)\rangle\,
  (g^{\mu\nu} - v^\mu v^\nu) \nonumber\\
&&\qquad = Y_n\, v_{\alpha_1} \ldots v_{\alpha_n} 
  + \mbox{(other tensor structures)}\,,
\eeqa
where the other tensor structures yield subleading contributions.  Collecting
all factors, the contribution to the lepton spectrum is
\beqa\label{4qspec}
{\d\Gamma\over \d y} &=& -{G_F^2 |V_{ub}|^2\, m_b^2 \over 3\, \pi}\,
  (g^{\mu\nu} - v^\mu v^\nu)\,
  \langle B(v)| (\bar b_v \gamma_\mu P_L u)\, \delta(1-y-i D \cdot n)\,
  (\bar u \gamma_\nu P_L b_v) |B(v)\rangle \nonumber\\*
&=& -{2G_F^2 |V_{ub}|^2\, m_b^2\over 3\pi}\,
  \sum_{n=0}^\infty {Y_n\over m_b^n\, n!}\, \delta^{(n)}(1-y) \,.
\eeqa
The first term in the expansion of the delta function reproduces Voloshin's
result in Eq.~(\ref{volo}), since  $Y_0 = (\langle B| O_{V-A} |B\rangle -
\langle B| O_{S-P} |B\rangle) /2 = (B_1-B_2) f_B^2\, m_B / 8$.

\begin{table}[t]
\begin{tabular}{@{\hspace{10pt}} c @{\hspace{10pt}} ||
  @{\hspace{10pt}} c @{\hspace{25pt}} c @{\hspace{20pt}} c @{\hspace{10pt}}}
  \hline\hline
Lepton energy  &  \multicolumn{3}{c@{\hspace{10pt}}}{Fractional correction, 
  $|\delta\Gamma|/\Gamma$} \\
intervals  &  $b = 1/2$  &  $b = 1$  &  $b = 3/2$ \\ \hline\hline
$2.0\,\mbox{GeV} < E_\ell$  &  9\%   &  9\%  &  8\%  \\
$2.1\,\mbox{GeV} < E_\ell$  &  12\%   &  12\%  &  10\% \\
$2.2\,\mbox{GeV} < E_\ell$  &  17\%   &  16\%  &  12\% \\
$2.3\,\mbox{GeV} < E_\ell$  &  28\%   &  23\%  &  15\% \\
$2.4\,\mbox{GeV} < E_\ell$  &  57\%   &  37\%  &  18\% \\ \hline\hline
\end{tabular}
\caption{Fractional corrections to the rate in the endpoint region due to the
smeared four-quark operators.  The first moment of this subleading structure
function need not vanish, so it is modelled by Eq.~(\ref{shapemodel}) with
$\bar\Lambda = 0.5\,$GeV, $a=2.5$ and the different values of $b$ shown in the
Table.  For the leading structure function we used $b=1$, $\bar\Lambda =
0.5\,$GeV, $a = 2.5$ (corresponding to $\lambda_1 = -0.3\, \mbox{GeV}^2$).}
\label{tab4q}
\end{table}

An important difference between the structure function for the four-quark
operators and those discussed explicitly in Secs.~II and III is that in the
present case the first moment of $f(k_+)$ need not vanish.  The importance of
this difference is illustrated in Table~\ref{tab4q}.  To obtain these estimates
we held the leading order structure function fixed, as discussed previously. 
For the subleading structure function we use Eq.~(\ref{shapemodel}) with
$\bar\Lambda = 0.5\,$GeV and $a = 2.5$ held fixed, while varying the $b$
parameter which determines the first moment.  Since such contributions do not
occur in radiative $B$ decay, using the $B\to X_s\gamma$ photon spectrum as an
input does not reduce these corrections.  The sign of this correction is
undetermined, and the magnitude is directly proportional to the assumed
violation of factorization (we used $|B_1 - B_2| = 0.1$ and $f_B = 200\,$MeV). 
It may be possible to address the size of factorization violation using lattice
QCD~\cite{MDP}.

\section{Conclusions}

We investigated subleading structure function contributions to $B\to
X_u\ell\bar\nu$ that are important in the region $E_\ell > 2\,$GeV, and could
each alter significantly the relation between the lepton spectrum and the $B\to
X_s\gamma$  photon spectrum used for the extraction of $|V_{ub}|$.  We
concentrated on two subleading structure functions.

The first series of corrections, considered in Sec.~III, is related to the 
smearing of the part of the $(\lambda_2/m_b^2)\, \delta(1-y)$ contributions to
the $B\to X_u\ell\bar\nu$ lepton spectrum originating from the one-gluon matrix
element in Fig.~1.  This is formally an order $\lqcd/m_b$ correction in the
endpoint region, however it enters with an enhanced coefficient, 6.  Its
contribution to the total $B\to X_u\ell\bar\nu$ rate is known to be $-3\%$, and
estimates of how it affects the lepton spectrum integrated over different
intervals near the endpoint are shown in Table~\ref{tab1G}.

The second series of corrections, considered in Sec.~IV, is related to the
smearing of four-quark operator contributions.  This is formally an order
$\lqcd^2/m_b^2$ correction in the endpoint region, however it is enhanced by
$16\pi^2$.  Contrary to the previous case, the sign of this contribution is
unknown, and its magnitude is also uncertain.  It could be a $\sim3\%$
contribution to the total $B\to X_u\ell\bar\nu$ rate, and estimates of how it
affects the lepton spectrum integrated over different intervals near the
endpoint are shown in Table~\ref{tab4q}.

These corrections could each affect the value of $|V_{ub}|$ extracted from the
lepton endpoint analysis above the 10\% level, even when it is combined with
the measurement of the $B\to X_s\gamma$ photon spectrum.   While varying the
lower end of the energy interval between 2\,GeV and 2.4\,GeV, as done in the
recent CLEO analysis, provides some indication that these corrections are
probably not much larger than our estimates, the uncertainty is reliably
reduced only if the lower limit of the lepton energy interval is near 2\,GeV,
in which case the lepton energy cut no longer eliminates the charm background. 
However, if the $B\to X_u\ell\bar\nu$ lepton spectrum can be measured
accurately down to 2\,GeV, then it is likely that a resummation of singular
terms in the OPE in the endpoint region is not necessary.  In that case,
knowledge of the $B\to X_s\gamma$ photon spectrum is not needed\footnote{The
first moment of the $B\to X_s\gamma$ photon spectrum may still be the best
observable to extract the $b$ quark mass from~\cite{bsgmass}, which is crucial
for any determination of $|V_{ub}|$ from inclusive $B$ decays.} and $|V_{ub}|$
can be extracted without a resummation of the leading singular terms.  In the
$E_\ell \gtrsim 2.2\,$GeV region, where the resummation of the leading
singularities in the OPE is important, so is the uncertainty related to the
subleading structure functions discussed in this paper.  With more precise
$B\to X_u\ell\bar\nu$ and $B\to X_s\gamma$ data in the region above 2.2\,GeV,
it may be possible to observe these corrections.

\acknowledgments

We thank Ed Thorndike for helpful discussions.
A.K.L.~would like to thank the LBL theory group for its hospitality
while portions of this work were done.  A.K.L.~was supported in part
by the Department of Energy under Grant DE-AC02-76CH03000.  Z.L.~was
supported in part by the Director, Office of Science, Office of High
Energy and Nuclear Physics, Division of High Energy Physics, of the
U.S.\ Department of Energy under Contract DE-AC03-76SF00098.
M.B.W.~was supported in part by the Department of Energy under Grant
No.~DE-FG03-92-ER40701.

\end{document}